\documentclass[conference]{IEEEtran}
\IEEEoverridecommandlockouts

\usepackage{amsmath,amssymb}
\usepackage{graphicx}
\usepackage{booktabs}
\usepackage{cite}
\usepackage{url}
\usepackage{enumitem}
\usepackage{float}
\usepackage{makecell}

\begin{document}

\title{\vspace{-0.6cm}Infrared Hotspot-Guided Early Warning of Lithium-Ion Battery Thermal Runaway Under Mechanical Abuse\vspace{-0.3cm}}

\author{
\IEEEauthorblockN{
\begin{tabular}{ccc}
\begin{minipage}[t]{0.30\textwidth}
\centering
\small Syed Sajid Ullah*\\[-1pt]
\scriptsize
\textit{School of Energy and Electrical Engineering}\\[-1pt]
\textit{Chang'an University}\\[-1pt]
Xi'an, 710064, China\\[-1pt]
sajid@chd.edu.cn
\end{minipage}
&
\begin{minipage}[t]{0.30\textwidth}
\centering
\small Salman Khan\\[-1pt]
\scriptsize
\textit{School of Energy and Electrical Engineering}\\[-1pt]
\textit{Chang'an University}\\[-1pt]
Xi'an, China
\end{minipage}
&
\begin{minipage}[t]{0.30\textwidth}
\centering
\small Muhammad Zunair Zamir\\[-1pt]
\scriptsize
\textit{School of Information Engineering}\\[-1pt]
\textit{Chang'an University}\\[-1pt]
Xi'an, 710064, China
\end{minipage}
\end{tabular}
}
}

\maketitle

\begingroup
\renewcommand\thefootnote{}
\footnotetext{*Corresponding author: Syed Sajid Ullah (sajid@chd.edu.cn).}
\endgroup

\begin{abstract}
Mechanical abuse can trigger thermal runaway (TR) in lithium-ion batteries through localized heat generation before sensor signals become decisive. This paper proposes a two-stage early-warning approach that estimates localized thermal instability from infrared hotspot dynamics and then fuses this instability score with mechanical, electrical, thermal, and image-intensity features for a 20-frame warning horizon. Evaluation uses repeated experiment-wise three-fold validation, with out-of-fold Stage-I scores during Stage-II training to prevent stacked-model optimism. Hotspot dynamics alone achieve Stage-I ROC-AUC 0.945, and the two-stage classifier reaches Stage-II ROC-AUC 0.908, exceeding direct multimodal fusion while preserving an interpretable intermediate instability signal. Thermal gradient rise precedes voltage-based detection by 40 frames (4 seconds) on average, enabling earlier battery management system intervention. Lead-time analysis at a fixed 0.5 threshold yields a 14.8-frame mean lead time.
\end{abstract}


\section{Introduction}

Lithium-ion batteries power electric vehicles, portable electronics, and grid-scale storage because of their high energy density and long cycle life. Safety failures remain a critical concern, especially when a cell undergoes thermal runaway (TR): rapid heat release, gas venting, fire, and propagation to adjacent cells~\cite{feng2018thermal}.

Mechanical abuse constitutes a major failure trigger~\cite{li2023multifield}. Indentation, compression, penetration, and crash-induced deformation damage internal layers, cause separator failure, and initiate internal short circuits. The resulting localized Joule heating can evolve into TR if undetected~\cite{liu2020safety}.

Conventional warning strategies depend on voltage drop, scalar surface temperature, force, or deformation thresholds~\cite{liu2024review}. These signals describe abuse progression, but they may not capture the spatially localized thermal response that precedes runaway. Infrared thermal imaging provides this spatial information through hotspot formation, growth, thermal gradient, entropy, and centroid motion~\cite{shan2024insights}. Existing work still leaves three practical gaps for compact early-warning pipelines: single-stage fusion can submerge thermal patterns among other features, temporal splits can leak frame-level correlations into test sets, and diagnostic reporting is often not separated cleanly from the prediction pipeline.

We present a two-stage early-warning strategy that decouples localized thermal instability detection from final warning prediction. Stage I estimates an annotated thermal-instability state from hotspot features alone. Stage II then combines this instability signal with mechanical, electrical, scalar thermal, and four global IR intensity statistics. Hotspot dynamics are therefore distilled into a physically interpretable instability score rather than being submerged directly among many multimodal inputs. Surface temperature can rise by over 150$^\circ$C during the stable stage, well before voltage collapse becomes detectable. This temporal gap creates an early window for intervention that conventional voltage-threshold methods cannot exploit. At the 10 Hz IR frame rate, the observed 40-frame thermal gradient lead corresponds to 4 seconds of advance warning before voltage-based detection, sufficient for a battery management system to trigger mitigation.

The core contributions are:
\begin{itemize}[nosep]
  \item A leakage-controlled two-stage architecture that distills thermal-image features into an intermediate instability score before multimodal fusion.
  \item A compact hotspot-dynamics representation for Stage-I instability estimation, achieving ROC-AUC 0.945 under experiment-wise validation.
  \item Structured diagnostic reporting that exposes classifier evidence while preserving the intermediate instability score as an interpretable monitoring variable.
\end{itemize}
The two-stage design targets interpretability and leakage-controlled evaluation while improving operating-threshold performance over direct multimodal fusion.

\section{Related Work}

Mechanically induced TR has been studied under indentation, compression, nail penetration, and impact~\cite{hu2025mechanistic}. These events induce internal short circuits and localized Joule heating, making early diagnosis challenging because the critical precursor can be spatially localized before global cell temperature rises~\cite{e2024comprehensive,xiao2024review}.

Sensor-based warning methods commonly use voltage, temperature, force, strain, gas, pressure, or ultrasonic measurements~\cite{wang2024monitoring,tam2025development,lee2025advanced}. These are practical and compatible with battery management systems, but scalar measurements may provide delayed or spatially incomplete information. Data-driven classifiers improve on fixed thresholds by learning nonlinear relationships among sensor signals, yet they can still underuse spatial thermal patterns~\cite{liu2024lithium}.

Thermal imaging captures hotspot formation, spatial thermal gradients, and propagation behavior~\cite{lin2025mechanically}. Recent multimodal studies have combined thermal images with sensor streams and interpretable machine-learning tools for thermal-runaway prediction~\cite{trunsafe2025,ml2025explainable}. Image and sensor fusion is a promising direction, but these studies do not always separate localized thermal-instability estimation from the final warning decision.

Language-model-assisted safety reporting relates to this goal but is not treated here as an independent predictor. Prior work on self-decoupled or multimodal sensing emphasizes the value of physically separated evidence channels for early warning~\cite{selfdecouple2025}. Our diagnostic layer follows this spirit, restricting itself to structured reporting from classifier outputs rather than free-form language classification.

Across these directions, prior studies have explored thermal imaging, multimodal fusion, and diagnostic interpretation separately. This work combines hotspot-derived instability scoring, experiment-wise leakage control, and structured diagnostic reporting within one compact early-warning pipeline. Table~\ref{tab:related_work} positions this work against representative studies in each direction.

\begin{table*}[ht]
\centering
\caption{Positioning against representative prior work.}
\label{tab:related_work}
\footnotesize
\begin{tabular}{lccccc}
\toprule
Reference & Mech. abuse & Hotspot dyn. & Fusion & Exp. split & Diag. reporting \\
\midrule
\cite{liu2024review} & Partial & No & No & N/A & No \\
\cite{wang2024monitoring,tam2025development,liu2024lithium} & Yes & No & Partial & Partial & No \\
\cite{lin2025mechanically} & Partial & Partial & Partial & Partial & No \\
\cite{trunsafe2025,ml2025explainable} & Partial & Partial & Yes & Partial & No \\
\cite{he2024reduced} & Partial & No & Partial & N/A & Yes \\
This work & Yes & Yes & Yes & Yes & Yes \\
\bottomrule
\end{tabular}
\end{table*}

\section{Methodology}

\subsection{Experimental Setup and Dataset}
\label{sec:dataset}

\begin{figure}[ht]
\centering
\includegraphics[width=\columnwidth]{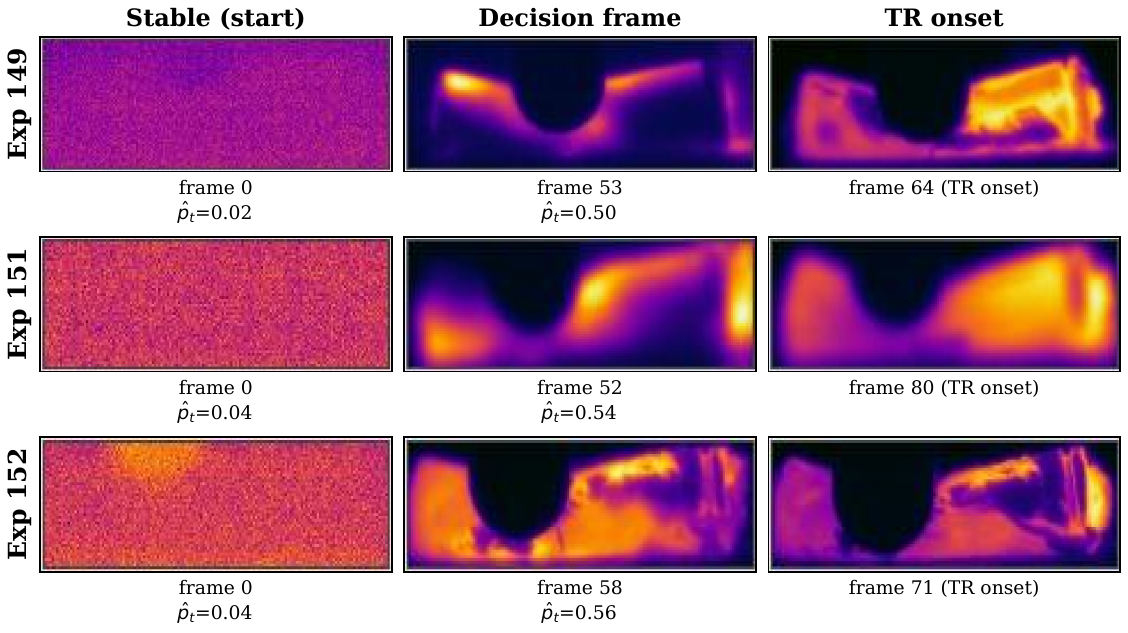}
\caption{Representative thermal IR frames for Experiments~149, 151, and~152 at the stable starting frame, the model's decision frame (with predicted warning probability $\hat{p}_t$), and the thermal-runaway (TR) onset frame, illustrating how the proposed two-stage model's warning signal tracks the visible thermal progression.}
\label{fig:thermal_grid}
\end{figure}

Cylindrical lithium-ion cells under indentation loading yield synchronized frame-level records~\cite{li2026multimodal} with mechanical, electrical, scalar thermal, and infrared thermal-image metadata at 10~Hz. Hotspot blobs extracted from IR images provide spatial dynamics (area, gradient, entropy, centroid coordinates, growth, and velocity). Runaway-propagation frames are excluded to concentrate on pre-runaway prediction, leaving 12,425 training rows from 194 experiments; five propagation-only experiments are retained only for threshold-based signal comparisons. Fig.~\ref{fig:thermal_grid} shows representative IR frames together with the proposed two-stage model's predicted warning probability $\hat{p}_t$ at key frames. Table~\ref{tab:dataset_summary} summarizes the dataset.

\begin{table}[ht]
\centering
\small
\caption{Dataset summary used for preliminary experiments.}
\label{tab:dataset_summary}
\resizebox{\columnwidth}{!}{%
\begin{tabular}{lc}
\toprule
Item & Value \\
\midrule
Frame-level observations & 40,730 \\
Number of experiments & 199 \\
Number of columns & 38 \\
Pre-runaway rows & 12,425 \\
Experiments used for training & 194 \\
Stage-I target & Stable vs. precursor/localized \\
Stage-II target & TR within next 20 frames \\
\bottomrule
\end{tabular}%
}
\end{table}

Feature groups (Table~\ref{tab:feature_groups}) organize inputs by physical role: mechanical, electrical, scalar thermal, image intensity, and hotspot dynamics.

\begin{table}[ht]
\centering
\small
\caption{Feature groups and their physical interpretation.}
\label{tab:feature_groups}
\resizebox{\columnwidth}{!}{%
\begin{tabular}{lcc}
\toprule
Group & Example features & Physical meaning \\
\midrule
Mechanical & force, deformation, dF/dt & Abuse severity \\
Electrical & voltage & Electrical response \\
Scalar thermal & high\_temp, low\_temp, temp\_range & Global surface heating \\
Thermal image & mean, std, max, min intensity & Image-level thermal state \\
Hotspot dynamics & area, centroid, gradient, entropy, growth, velocity & Localized thermal instability \\
\bottomrule
\end{tabular}%
}
\end{table}

The Stage-I target merges precursor and localized-instability frames into one positive instability class against stable frames; Stage II uses a binary label marking whether TR occurs within the next 20 frames. Event-derived variables are excluded to prevent leakage.

\subsection{Problem Formulation}
\label{sec:formulation}

For each frame $t$, the multimodal feature vector is
\begin{equation}
\mathbf{x}_t = [\mathbf{x}^{m}_t, \mathbf{x}^{e}_t, \mathbf{x}^{th}_t, \mathbf{x}^{int}_t, \mathbf{x}^{hot}_t],
\label{eq:feature_vector}
\end{equation}
where $\mathbf{x}^{m}_t$, $\mathbf{x}^{e}_t$, $\mathbf{x}^{th}_t$, $\mathbf{x}^{int}_t$, and $\mathbf{x}^{hot}_t$ denote mechanical, electrical, scalar thermal, image-intensity, and hotspot-dynamics features.

In Stage I, the localized thermal-instability state is estimated from hotspot-dynamics features only:
\begin{equation}
\hat{s}_t = P(z_t=1 \mid \mathbf{x}^{hot}_t),
\label{eq:stage1}
\end{equation}
where $z_t$ is the annotated thermal-instability label and $\hat{s}_t \in [0,1]$ is the estimated instability score. Restricting Stage I to hotspot dynamics ensures $\hat{s}_t$ depends purely on observed thermal spatial patterns.

In Stage II, the instability score augments compact sensor and image-intensity features:
\begin{equation}
\tilde{\mathbf{x}}_t = [\mathbf{x}^{m}_t, \mathbf{x}^{e}_t, \mathbf{x}^{th}_t, \mathbf{x}^{int}_t, \hat{s}_t],
\label{eq:augmented}
\end{equation}
and the early-warning probability is estimated as
\begin{equation}
\hat{p}_t = P(y_t=1 \mid \tilde{\mathbf{x}}_t),
\label{eq:stage2}
\end{equation}
where $y_t$ indicates whether TR occurs within the next 20 frames.

\subsection{Baseline Classifiers}
\label{sec:baselines}

\begin{itemize}[nosep]
  \item \textbf{Sensor-only}: Mechanical, electrical, and scalar-thermal measurements.
  \item \textbf{Image-all}: Raw intensity statistics plus hotspot-dynamics features.
  \item \textbf{Hotspot-only}: Area, centroid, gradient, entropy, growth-rate, and velocity features.
  \item \textbf{Direct multimodal}: All feature groups fused in one stage.
  \item \textbf{Proposed two-stage}: Stage I produces $\hat{s}_t$, and Stage II combines it with sensor features and four raw intensity statistics.
\end{itemize}

Stage I compares Sensor-only, Image-all, Hotspot-only, and Multimodal; Stage II compares Sensor-only, Image-all, Direct multimodal, and the proposed two-stage design under the same grouped folds. Deep-learning baselines (LSTM, CNN-LSTM, MLP) use 10-frame input sequences, Adam at $10^{-3}$, batch size 128, up to 30 epochs, and patience 6. Ablations remove the Stage-I score, raw intensity statistics, or time from the proposed configuration.

\subsection{Two-Stage Architecture}
\label{sec:proposed}

\begin{figure*}[ht]
\centering
\includegraphics[width=\textwidth]{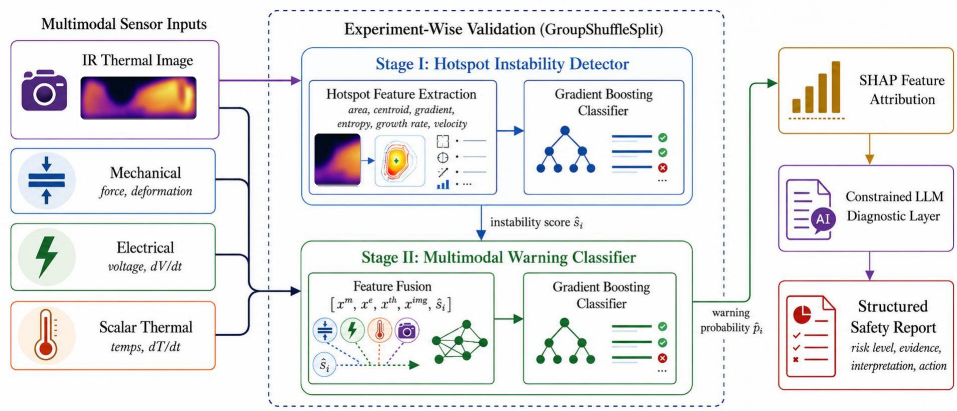}
\caption{Two-stage early-warning architecture. Stage I maps hotspot dynamics to an instability score, Stage II fuses that score with sensor and image-intensity features for 20-frame-ahead TR warning, and the diagnostic layer reports structured evidence without altering the classifier output.}
\label{fig:architecture}
\end{figure*}

Fig.~\ref{fig:architecture} illustrates the architecture, with two sequential stages and a constrained diagnostic reporting layer. Stage I receives hotspot dynamics only and outputs an instability probability. Stage II does not re-feed those hotspot descriptors directly; instead, it fuses the out-of-fold instability score with sensor variables and four global image-intensity statistics, creating an interpretable bottleneck between localized thermal evidence and the final warning decision. The diagnostic layer then reports structured evidence from the trained classifier without changing the prediction itself.

A LightGBM gradient-boosting classifier is used for both stages because the task is low-dimensional tabular fusion with nonlinear cross-modality interactions and occasional missing hotspot coordinates, for which tree boosting is stronger and easier to audit than a larger end-to-end network. All LightGBM variants share the same hyperparameters and grouped folds for a fair comparison. During training, Stage II receives out-of-fold Stage-I scores for training rows; the Stage-I classifier is then refitted on the outer training fold to score the outer test fold, preventing overfitted in-sample predictions. SHapley Additive exPlanations (SHAP) are then computed on the trained models to inspect which variables dominate each stage.

\subsection{Evaluation Metrics}
\label{sec:metrics}

All classifiers use experiment-wise three-fold cross-validation repeated over three random seeds~\cite{jeong2024prediction}. Table~\ref{tab:metrics} summarizes the evaluation metrics.

\begin{table}[H]
\centering
\footnotesize
\caption{Evaluation metrics.}
\label{tab:metrics}
\begin{tabular}{lll}
\toprule
Metric & Stage & Description \\
\midrule
ROC-AUC & I, II & Area under ROC curve \\
AP  & I, II & Average precision \\
Lead time & II & Frames from first warning to TR onset \\
\bottomrule
\end{tabular}
\end{table}

Mean ROC-AUC with min-max range across folds is reported for both stages. Tables also report AP, F1, and balanced accuracy (BAcc) at the default threshold of 0.5. Lead time is evaluated at the same threshold; a valid warning must cross the threshold before TR and within the 20-frame horizon.

\section{Results and Discussion}

\begin{figure}[H]
\centering
\includegraphics[width=0.48\columnwidth]{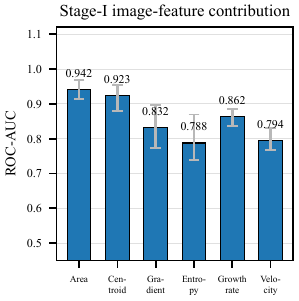}
\hfill
\includegraphics[width=0.48\columnwidth]{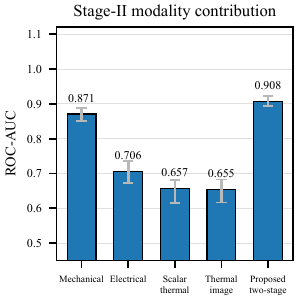}
\caption{(a) Stage-I hotspot-feature contribution to instability detection. (b) Stage-II single-modality contribution to early warning, with the proposed two-stage model shown for comparison.}
\label{fig:auc_grid}
\end{figure}

This section reports model performance for both stages of the two-stage framework, ablation results, and diagnostic feature analysis. Fig.~\ref{fig:auc_grid} summarizes the Stage-I hotspot-feature contributions and the Stage-II single-modality contributions.

\subsection{Stage I: Thermal Instability Detection}

\begin{figure}[ht]
\centering
\includegraphics[width=\columnwidth]{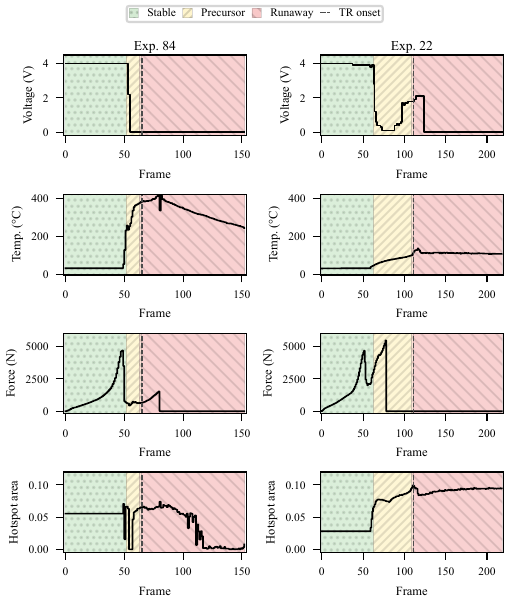}
\caption{Multimodal sensor trajectories for Experiments~84 and~22. Stage backgrounds combine color with light gray hatch patterns for black-and-white readability: green (stable), yellow (precursor), orange (localized), red (runaway). Dashed line: TR onset. Thermal signals become abnormal before voltage reaches a critical reading.}
\label{fig:sensor_trajectories}
\end{figure}

Fig.~\ref{fig:sensor_trajectories} shows why thermal imaging provides earlier diagnostic information than conventional signals. In Experiment~84, surface temperature rises from 31$^\circ$C to 230$^\circ$C during the stable stage while voltage remains at 4.0~V and hotspot area shows no appreciable growth. Voltage collapses to 0~V only in the precursor stage, by which point temperature has already exceeded 256$^\circ$C. In Experiment~22, a more gradual failure mode appears: voltage declines from 4.0~V to 1.8~V while hotspot area grows consistently. In both cases, thermal signals reveal anomaly progression before voltage reaches a decisively abnormal reading, motivating a pipeline that converts hotspot dynamics into an explicit instability score before final fusion.

Table~\ref{tab:thermal_early_advantage} quantifies this temporal advantage across all 199 experiments using threshold-based detection. Thermal gradient rise provides the earliest warning with a mean lead of 40.3 frames before runaway, compared to 21.0 frames for voltage drop below 3.5~V. It detects pre-runaway anomaly in 143 of 199 experiments (71.9\%) versus only 74 for the voltage threshold, confirming broader but not universal coverage.

\begin{table}[ht]
\centering
\footnotesize
\caption{Frame-level lead time before thermal runaway for threshold-based detection.}
\label{tab:thermal_early_advantage}
\resizebox{\columnwidth}{!}{%
\begin{tabular}{llllll}
\toprule
Signal & \makecell{Mean lead\\(frames)} & \makecell{Median\\lead} & \makecell{Min\\lead} & \makecell{Max\\lead} & \makecell{Exp.\\detected} \\
\midrule
Voltage $<$ 3.5 V & 21.0 & 13 & 1 & 125 & 74 \\
Surface temp. $>$ 50$^\circ$C & 21.1 & 13 & 1 & 115 & 76 \\
Hotspot area $>$ 1.5$\times$ baseline & 21.9 & 13 & 1 & 96 & 44 \\
Thermal gradient rise & 40.3 & 42 & 1 & 171 & 143 \\
\bottomrule
\end{tabular}%
}
\end{table}

Stage-I results appear in Table~\ref{tab:stage1_results}, while Fig.~\ref{fig:auc_grid}a disaggregates the hotspot branch into six feature families. Area-based cues are strongest (ROC-AUC 0.942), followed by centroid features (0.923); growth rate (0.862) and gradient (0.832) remain useful but weaker, while entropy (0.788) and velocity (0.794) are the least discriminative on their own. Sensor-only achieves ROC-AUC 0.881, reflecting limited early-warning information from scalar measurements alone. Hotspot-only achieves 0.945, confirming that the combined hotspot descriptor set closely matches annotated precursor and localized-instability states. Image-all performs similarly (0.946), while Multimodal achieves the highest ROC-AUC of 0.949, indicating complementary information between hotspot and sensor features.

\begin{table}[ht]
\centering
\small
\caption{Stage-I thermal instability detection results across repeated experiment-wise folds.}
\label{tab:stage1_results}
\resizebox{\columnwidth}{!}{%
\begin{tabular}{lcccccc}
\toprule
Model & ROC-AUC mean & ROC-AUC min & ROC-AUC max & AP mean & F1 mean & BAcc mean \\
\midrule
Sensor-only & 0.881 & 0.833 & 0.934 & 0.747 & 0.688 & 0.840 \\
Sensor-no-time & 0.880 & 0.828 & 0.934 & 0.752 & 0.688 & 0.840 \\
Image-all & 0.946 & 0.930 & 0.966 & 0.753 & 0.726 & 0.864 \\
Hotspot-only & 0.945 & 0.931 & 0.958 & 0.732 & 0.726 & 0.866 \\
Multimodal & 0.949 & 0.929 & 0.974 & 0.783 & 0.725 & 0.855 \\
\bottomrule
\end{tabular}%
}
\end{table}

\subsection{Stage II: Early-Warning Prediction}

Table~\ref{tab:model_comparison} summarizes Stage-II performance together with the deep-learning baselines, and Fig.~\ref{fig:auc_grid}b isolates the predictive strength of each physical modality when used alone. Mechanical variables are strongest (ROC-AUC 0.871), ahead of electrical (0.706), scalar thermal (0.657), and the thermal-image pathway (0.655) -- all well below the proposed two-stage model's 0.908, also plotted in Fig.~\ref{fig:auc_grid}b for direct comparison. The full two-stage classifier reaches ROC-AUC 0.908 versus 0.903 for Direct multimodal fusion, with stronger threshold-level gains (F1 0.752 vs.~0.738; BAcc 0.836 vs.~0.824). LightGBM with the two-stage design also outperforms all three deep-learning baselines by 5 to 7 ROC-AUC points.

\begin{table}[H]
\centering
\footnotesize
\caption{Stage-II early-warning comparison: LightGBM and deep-learning baselines.}
\label{tab:model_comparison}
\resizebox{\columnwidth}{!}{%
\begin{tabular}{lcccccc}
\toprule
Model & \makecell{ROC-AUC\\mean} & \makecell{ROC-AUC\\min} & \makecell{ROC-AUC\\max} & AP & F1 & BAcc \\
\midrule
\multicolumn{7}{l}{\textit{LightGBM variants}} \\
\quad Sensor-only & 0.896 & 0.881 & 0.916 & 0.781 & 0.733 & 0.821 \\
\quad Image-all & 0.683 & 0.666 & 0.706 & 0.572 & 0.498 & 0.637 \\
\quad Direct multimodal & 0.903 & 0.884 & 0.923 & 0.795 & 0.738 & 0.824 \\
\quad Proposed two-stage & \textbf{0.908} & 0.894 & 0.924 & \textbf{0.796} & \textbf{0.752} & \textbf{0.836} \\
\midrule
\multicolumn{7}{l}{\textit{Deep-learning baselines (sequence length 10)}} \\
\quad LSTM & 0.850 & 0.838 & 0.860 & 0.731 & 0.691 & 0.764 \\
\quad CNN-LSTM & 0.843 & 0.825 & 0.857 & 0.719 & 0.687 & 0.760 \\
\quad MLP & 0.839 & 0.827 & 0.861 & 0.684 & 0.694 & 0.767 \\
\bottomrule
\end{tabular}%
}
\end{table}

\begin{figure}[ht]
\centering
\includegraphics[width=\columnwidth]{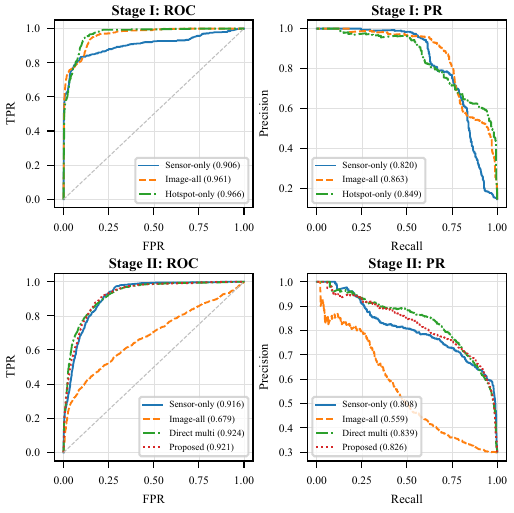}
\caption{ROC and precision-recall curves for Stage I and Stage II classifier comparisons. AUC and AP values appear in the legend.}
\label{fig:roc_pr_grid}
\end{figure}

Full ROC and precision-recall curves for both stages appear in Fig.~\ref{fig:roc_pr_grid}.

\subsection{Ablation Study and Diagnostic Reporting}

Ablation results appear in Table~\ref{tab:ablation_results}. Removing the Stage-I score reduces ROC-AUC from 0.908 to 0.905, while removing the raw intensity statistics reduces it further to 0.896. The strongest Stage-II behavior therefore comes from combining the explicit instability score with a small amount of global image context rather than feeding hotspot dynamics directly into the warning model.

\begin{table}[ht]
\centering
\small
\caption{Ablation study for the Stage-II early-warning model.}
\label{tab:ablation_results}
\resizebox{\columnwidth}{!}{%
\begin{tabular}{lcccc}
\toprule
Model & ROC-AUC mean & AP mean & F1 mean & BAcc mean \\
\midrule
Direct multimodal & 0.903 & 0.795 & 0.738 & 0.824 \\
Proposed two-stage & 0.908 & 0.796 & 0.752 & 0.836 \\
Ablation: no Stage-I score & 0.905 & 0.794 & 0.751 & 0.835 \\
Ablation: no intensity stats & 0.896 & 0.781 & 0.733 & 0.821 \\
Ablation: no time & 0.902 & 0.791 & 0.743 & 0.828 \\
\bottomrule
\end{tabular}%
}
\end{table}

\begin{figure}[ht]
\centering
\includegraphics[width=\columnwidth]{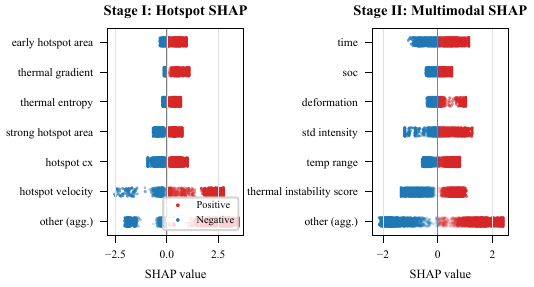}
\caption{SHAP summary plots for Stage~I (hotspot-only) and Stage~II (two-stage). Red: positive impact toward instability or warning; blue: negative impact. Each panel shows the top 6 features plus an aggregated ``Other'' row for the remaining contributions.}
\label{fig:shap_grid}
\end{figure}

SHAP analysis (Fig.~\ref{fig:shap_grid}) provides feature-attribution insight beyond raw importance scores. In Stage I, the top hotspot features (growth rate, thermal gradient, area) show asymmetric SHAP distributions: positive values increase the instability score, confirming that growing, spatially non-uniform heating drives the classifier toward instability detection. In Stage II, the injected instability score is the dominant feature, followed by compact image-intensity and voltage-related cues, indicating that the final warning combines localized hotspot evidence with coarse global heating and electrical response. An aggregated ``Other'' row preserves the remaining lower-magnitude SHAP contributions.

\begin{figure}[ht]
\centering
\includegraphics[width=\columnwidth]{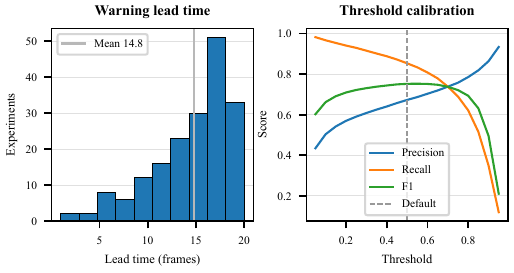}
\caption{(left) Warning lead-time distribution. (right) Precision, recall, and F1 as functions of decision threshold for Stage~II. The default 0.5 threshold is near-optimal for F1.}
\label{fig:timing_grid}
\end{figure}

At threshold 0.5, the model produces 183 valid in-horizon warnings with a mean lead time of 14.8 frames, and the remaining detections occur beyond the 20-frame horizon, indicating early risk awareness. The best mean F1 occurs at 0.55, but its gain over 0.5 is below 0.001, so we keep 0.5 as the default and leave stricter precision-recall trade-offs to deployment-specific tuning.

These results remain dataset-specific: the learning-based evaluation uses 194 experiments with pre-runaway frames, and the earliest handcrafted cue appears in 143 of 199 experiments rather than all cases. Transfer to other cell formats, abuse modes, chemistries, or pack-level propagation scenarios remains unvalidated. The diagnostic reporting module only verbalizes structured evidence from the trained warning model and does not alter the classifier decision~\cite{he2024reduced}.

\section{Conclusion}

We presented a two-stage thermal hotspot-aware framework for early warning of mechanically induced lithium-ion battery TR. Thermal signatures (temperature, gradient, hotspot area) precede voltage collapse by up to 40 frames on average, motivating a decoupled design that first distills hotspot dynamics into a localized instability score (Stage~I ROC-AUC 0.945) before compact multimodal fusion (Stage~II ROC-AUC 0.908). SHAP analysis confirms that this instability score remains the dominant Stage-II driver.

Future work should validate the model across additional abuse modes, cell chemistries, and pack-level propagation settings while studying deployment-specific threshold calibration and end-to-end thermal-video alternatives.

\end{document}